\documentclass[final,5p,times,twocolumn]{elsarticle}

\usepackage{graphicx}
\usepackage{amssymb}
\usepackage{amsmath}
\usepackage{empheq} 
\usepackage{comment}
\usepackage{balance}
\usepackage{enumerate}
\usepackage[inline,shortlabels]{enumitem}

\usepackage{relsize}	

\usepackage[usenames,dvipsnames]{color}
\usepackage[	colorlinks=true,
			linkcolor=NavyBlue,
			citecolor=NavyBlue,
			urlcolor=NavyBlue]{hyperref} 

\biboptions{sort&compress} 

\makeatletter
\def\@author#1{\g@addto@macro\elsauthors{\normalsize%
    \def\baselinestretch{1}%
    \upshape\authorsep#1\unskip\textsuperscript{%
      \ifx\@fnmark\@empty\else\unskip\sep\@fnmark\let\sep=,\fi
      \ifx\@corref\@empty\else\unskip\sep\@corref\let\sep=,\fi
      }%
    \def\authorsep{\unskip,\space}%
    \global\let\@fnmark\@empty
    \global\let\@corref\@empty  
    \global\let\sep\@empty}%
    \@eadauthor={#1}
}
\makeatother

\makeatletter
\def\ps@pprintTitle{%
 \let\@oddhead\@empty 
 \let\@evenhead\@empty
 \def\@oddfoot{}%
 \let\@evenfoot\@oddfoot}
 \makeatother

\journal{~}
\begin{document}
\begin{frontmatter}

\title{A Theory of Discrete Hierarchies\\ as Optimal Cost-Adjusted Productivity Organisations}

\author[add1]{Sandro Claudio Lera}
\ead{slera@ethz.ch}

\author[add2]{Didier Sornette}
\ead{dsornette@ethz.ch}

\address[add1]{\scriptsize The Media Lab, Massachusetts Institute of Technology, 77 Massachusetts Avenue, 02139 Cambridge, Massachusetts, USA}
\address[add2]{\scriptsize ETH Zurich, Department of Management, Technology, and Economics, Scheuchzerstrasse 7, 8092 Zurich, Switzerland}

\begin{abstract}
Hierarchical structures are ubiquitous in human and animal societies, but a
fundamental understanding of their raison d'\^etre has been lacking.
Here, we present a general theory in which hierarchies are obtained as the optimal
design that strikes a balance between the benefits of group productivity and the costs of 
communication for coordination. 
By maximising a generic representation of the output of a hierarchical organization
with respect to its design,  the optimal configuration of group sizes at different levels can be determined. 
With very few ingredients, a wide variety of hierarchically ordered complex organisational structures can be derived. 
Furthermore, our results rationalise the ubiquitous occurrence of triadic hierarchies, i.e., of the 
universal preferred scaling ratio between $3$ and $4$ found in many human and animal hierarchies, which should occur 
according to our theory when 
production is rather evenly contributed by all levels.
We also provide a systematic approach for optimising team organisation, 
helping to address the question of the optimal `span of control'. The significantly larger number $\sim 3-20$ of 
subordinates a supervisor typically manages is rationalised to occur in organisations
where the production is essentially done at the bottom level 
and in which the higher levels are only present to optimise coordination and control.
\end{abstract}

\end{frontmatter}

Throughout most of Homo sapiens 300'000 year record, humans have lived in small-scale, mostly egalitarian hunter-gatherer societies, 
comprising around 30-50 or, at most, a few hundred individuals \cite{Dunbar1998,Shultzineretal2010,Zafeiris2017}. 
Following the strong warming of Earth by 5 to 10 $^\circ$C from about 15'000 years ago leading to the 
end of the last ice age, settled communities emerged around 10'000 years ago, together with agriculture and animal domestication. 
These societies have been mostly structured into hierarchical societies.
Over the past millennia, even more complex, large scale interconnected societies have evolved, shaped into cultural, economic, political and corporate hierarchies \cite{Zafeiris2017,Turchin2018}. 
Explanations for the benefits of hierarchical organisation are manifold, such as advantages in warfare and multilevel selection \cite{Gavrilets2010,Zafeiris2017},
optimal search properties \cite{Maieretal18}, robustness \cite{SmithPuzio}, effective use of resources \cite{Zafeiris2017} and so on. 
But a framework to quantitatively relate the specific hierarchical structures to the functions and constraints facing different types of society has been lacking.

Here, we determine the optimal social hierarchical configuration by maximising the output of an organization with respect to its design. Our framework accounts for the finite Dunbar's number as well as the universal preferred scaling ratio between $3$ and $4$ found in many human and animal hierarchies \cite{Zhou2005,Dunbar2015,Sutcliffe2016,Tamarit2018}. 
This model provides the first quantitative explanation for the ubiquitous occurrence of such triadic hierarchies, and furthermore provides a framework to answer questions regarding optimal team sizes in 
management tasks, helping to address the question of the optimal \textit{span of control} \cite{Entwisle1961,Tarng1988,Bell1967}.

\section{Ingredients of a reduced-form theory of organisation structure} 
\label{sec:intro}

\subsection{Production scaling and costs of coordination as a function of group size}

We consider $N$ individuals, who are working together to produce some output. 
This can be a directly measurable product or quantity, such as the revenue of a firm,
or a more abstract quantity, e.g. overall group fitness. 
Quite generally, we may assume that the joint production $\Pi$, resulting from the interaction of $N$ individuals, scales as
$\Pi \sim N^\beta ~~~ (\beta > 0).$
The most straightforward situation corresponds to $\beta =1$, i.e. global output is proportional to population.
However, for small groups, one could expect that ``the whole is more than the sum of its parts'', and indeed, 
it has been shown that the aggregate output in open-software projects scales super-linearly with the number of developers $(\beta > 1)$, at least for group sizes $N$ less than 30 to 50 persons \cite{Sornette2014c}.
Intuitively, this means that a group of individuals together can produce more than the sum of their individual production 
in absence of interaction. Generally, specialisation and complementary skills motivate 
cooperation between individuals to achieve results that would otherwise be impossible.
Increased productivity can result from information sharing \cite{Lachmann2000}
as well as group heterogeneity \cite{Sornette2014c}, among others.

But as the group size increases, this super-linear production may tip over to
just linear ($\beta = 1$) or even sub-linear growth ($\beta < 1$)  \cite{Scholtes2016,Maillart2016},
because the human brain can only cope with a limited number of social interactions \cite{Dunbar2007}
and too many communication channels would overload the attention span leading to collapsing performance.
Generally, the overhead associated with communication, coordination and management 
of a group of collaborators of size $N$ tends to decrease 
the performance per individual, a well-known characteristic of large organisations.
As a first step, this cost can be represented as being proportional to 
the number $N (N-1) /2 \sim N^2$ of pair-wise interactions between the $N$ individuals in the group.

\subsection{Optimal group size and Dunbar's number}
\label{sec:Dunbar_number}

Starting from the production scaling law $\Pi \sim N^\beta$ supposed to hold for small teams,
and adding communication costs as being approximately proportional to $N (N-1) /2 \sim N^2$, we obtain
\begin{equation}
	\Pi = \mu N^\beta - \lambda  N^2, 
	\label{eq:Pi_simple_scaling_communication}
\end{equation}
where $\mu$ and $\lambda$ are two positive constants, which we refer to as the
productivity factor and the coordination cost factor, respectively.
Production now exhibits a maximum at 
\begin{equation}
N^* = \left(  \mu \beta / 2 \lambda  \right)^{1/(2-\beta)}
\label{eq:Nopt} 
\end{equation}
 individuals (as long as $\beta < 2$, which is a realistic assumption \cite{Sornette2014c}).
Thus, rather than a production scaling with the group size $N$, expression \eqref{eq:Pi_simple_scaling_communication}
predicts that, due to the cost of communication and coordination,  groups of sizes larger than $N^*$ persons
produce less than smaller groups of size $N^*$. Large scale societies would then collapse into 
independent groups of size $N^*$. While this is obviously counterfactual when interpreted for production, this prediction provides
a rational for Dunbar's number \cite{Dunbar1992}, which is the maximum number of people with whom one can
and does maintain stable social relationships. Dunbar's number is typically between 100 and 250, with a commonly used 
typical value of $150$. This finite number has been suggested to result from cognitive constraints on group size
that depends on the volume of neural material available for processing and synthesizing information
on social relationships.  This `social brain hypothesis' describes the coevolution
of neocortical brain size and social group sizes.
In this context, the first term $\sim N^\beta$ captures the need for humans to cooperate and to socialise. The second term 
$\sim N^2$ embodies the costs of enforcing the restrictive rules and norms to maintain a stable, cohesive group.  
Using $\beta \approx 1.5$ \cite{Sornette2014c} in \eqref{eq:Nopt} yields $\mu / \lambda \approx 16$ for humans.
Within this simple framework, the smaller group sizes of monkeys and primates may be interpreted as 
due to a smaller productivity factor and/or a larger coordination cost factor. Evolutionary improvement
of the productivity factor by a factor of two predicts a four-fold increase of the optimal group size (for a fixed $\beta =1.5$),
possibly explaining how moderate cognitive increase may be associated with much larger group sizes.
Technology, in the form of digital networking and artificial intelligence for instance, might promote
an increase in the productivity factor, which could then be associated with larger social group sizes in futuristic
human-digital symbiotic societies.

\subsection{Evidence and needs for sub-group formation}

Returning to the description of large firms and countries, their overall outputs typically increase
approximately in proportion to the number of employees or citizens
(allowing to define for instance such important economic metric as the GDP per capita).
What is missing in the naive model \eqref{eq:Pi_simple_scaling_communication} is that, in a group of $N$ individuals, 
not everybody is directly interacting with everybody else. 
Instead, sub-groups form, with closely knitted individuals within a given sub-group interacting with other sub-groups 
via their representatives. A vivid illustration is provided by the organisation of combattants in an army, where soldiers
at the bottom level form squads of about 10 headed by a corporal, then 3-4 squads form a platoon, 3 platoons 
combine into a company and so on.  Such an organisation ensures an efficient transmission of information
top-down and bottom-up for optimal battlefield performance. Such tendency to arrange into hierarchically structured groups 
have been reported widely, as previously mentioned \cite{Zhou2005,Fuchs2014,Zafeiris2017}. 

To develop an intuition how this can come about, let us 
consider again $N$ agents who need to communicate and coordinate. Under a flat organisation
in which everyone interacts with everyone, the total coordination cost would be $C \sim  N^2$.
But dividing the population into $N_1$ groups of $N_0$ individuals each ($N=N_0 \cdot N_1$),
the total communication overhead $C$ then scales as $N_0^2 \cdot N_1 + N_1^2$, where the first term accounts for the
intra-communication cost of $N_1$ groups of size $N_0$, and the second term accounts for the inter-communication 
between the $N_1$ groups through a single channel (for instance one representative of each group).
$C$ is minimized for $N_0 \sim N^{1/3}, N_1 \sim N^{2/3}$ for which $C \sim N^{4/3}$. 
The introduction of an additional level structure above the individual one thus reduces 
the communication overhead very significantly from $C \sim N^2$ to $N^{4/3}$.
In the supplementary information (SI), building on \cite{Toulouse1978}, we show that the addition 
of more layers (groups of groups, and so on) asymptotically reduces the cost to $C \sim N$ (and some logarithmic correction terms). 
This reduced communication cost in hierarchical organizations helps understand how states and companies 
can function even when $N$ is of the order of millions. 

However, the argument that hierarchical structures are created just to solve the coordination problem \cite{Toulouse1978}
cannot be the whole story, because social agents come together in the first place to gain something, such 
as mutual protection, increased outputs, and so on. 
Here, we extend expression \eqref{eq:Pi_simple_scaling_communication} to general hierarchical organisation structures
and derive the optimal designs to maximise production.

\section{Parameterisation and formulation of the optimisation problem}
\label{sec:principle_max_production}

\begin{figure}
	\centering
	\includegraphics[width=0.5\textwidth]{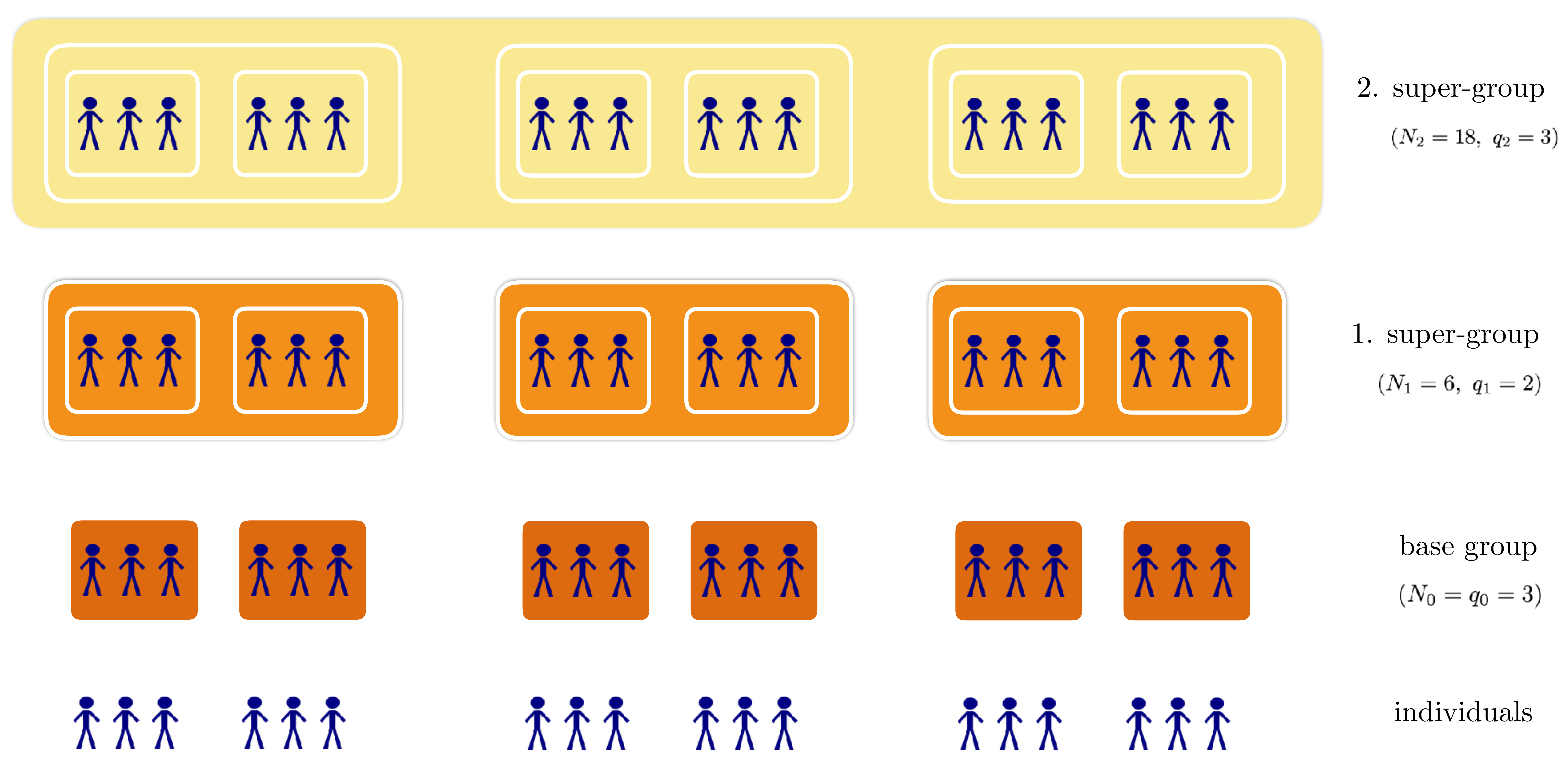}
	\caption{	Illustration of the hierarchical organization.
			We show the case with $p = 2$, i.e. $3$ hierarchical levels (without counting the individual level). 
			We start from $N = 18$ individuals. 
			The base groups are of size $3$ $(N_0 = q_0 = 3)$, i.e. three individuals together form one group. 
			The next groups (the first super-groups) are of size $2$ ($q_1 = 2, N_1 = q_1 \cdot q_0 = 6$), i.e. two base-groups together form one higher order group.
			The second super-group (and also the top level) is again of size $3$ ($q_2 = 3, N_2 = q_0 \cdot q_1 \cdot q_2 = N = 18$). 
			}
	\label{fig:illustration}
\end{figure}

\subsection{General formulation}

Following \cite{Toulouse1978}, we consider $N$ individuals organised  into $p$ hierarchical levels.
We denote by $N_0$ the number of individuals per group at the bottom of the hierarchical structure, i.e. the number of individuals per `base group'. 
At the next higher order in the hierarchical chain, $q_1 = N_1/N_0$ base groups taken together form a supergroup of $N_1$ individuals. 
Iterating, we define $q_r$ as the hierarchical group ratio, i.e. the size of a group at level $r$ compared to level $r-1$, 
\begin{equation}
	q_r \equiv \frac{N_r}{N_{r-1}} 
	= 
	\parbox{15em}{number of groups of size $N_{r-1}$ that \\ form a supergroup of $N_r$ individuals,}
	\label{eq:defi_q_r}
\end{equation}
with $q_0 \equiv N_0$, and deduce iteratively $N_r = q_0 \cdot  q_1 \cdot \ldots \cdot q_r$
Through $q_p = N_p / N_{p-1}$ and $N_p \equiv N$, we arrive finally at the highest level of the hierarchy. 

Fig. \ref{fig:illustration} illustrates this construction. 
Note in particular that because $N_p = N$ is the number of individuals at the highest level of the hierarchy, 
we have a total of $p+1$ hierarchical levels (we do not count the individual level). 
The special case of absence of hierarchies, i.e.
$p = 0, q_0 = N_0 = N$, represents a system with only one level
where everyone interacts with everyone. Identifying $N_{-1} \equiv 1$ allows us to treat this case consistently. 
Also, note that $q_r \geqslant 2$ (groups consist of at least two members). 
The maximum number of hierarchical levels is then $p_\text{max} = \lfloor \log_2 N \rfloor - 1$ (where 
$\lfloor x  \rfloor$ denotes the integer part of $x$), with the constant $-1$ ensuring that 
counting starts from $p=0$.

The hierarchical generalization of expression \eqref{eq:Pi_simple_scaling_communication} to $p+1$ hierarchical levels 
amounts to summing over the productions $\Pi_r$  of each level $r$ as follows:
\begin{equation}
	\Pi(p)
	= \sum_{r=0}^p \Pi_r 
	= \sum_{r=0}^p ~ \left(  \mu_r q_r^\beta - \lambda_r q_r (q_r-1)  \right) \times \frac{N}{\prod_{i=0}^r q_i}
	\label{eq:Pi_general_explicit}
\end{equation}
where the first factor under the sum in the r.h.s. of \eqref{eq:Pi_general_explicit} denotes the production of a group at level $r$ and the second factor represents the number of groups at that level. 

\subsection{Geometric hierarchies}

The productivity factors $\mu_r$ and coordination cost factors $\lambda_r$ can depend on level $r$.
It is natural to consider a {\it geometric hierarchy} defined with 
\begin{equation}
\mu_r= \omega~ \kappa^r,~~~~\lambda_r = \rho^r~, 
\label{eq:coef_definition}
\end{equation}
for some positive numbers $\omega, \kappa, \rho$. The geometric series for $\mu_r$ and $\lambda_r$
with constant scaling factors $\kappa$ and $\rho$ give a parsimonious dependence on the level $r$:
for $\kappa>1$ (resp. $<1$), higher levels of the hierarchy are more (resp. less) productive;
for $\rho >1$ (resp. $<1$), higher levels of the hierarchy require more (resp. less) efforts for coordination.
The special case $\kappa=1$ (resp. $\rho =1$) corresponds to the same productivity (resp. communication cost)
at all levels. The additional coefficient $\omega$ in $\mu_r$ is the production of a single individual,
which also sets the relative strength of productivity versus communication cost.  

Putting \eqref{eq:coef_definition} into \eqref{eq:Pi_general_explicit}, factoring out $\kappa^r$ and defining 
the relative cost-productivity scaling factor $\eta \equiv \rho / \kappa$ gives
\begin{equation}
	\Pi(p) = 
	 \sum_{r=0}^p ~\kappa^r~ \left( \omega~q_r^\beta -  \eta^r ~q_r (q_r-1)  \right) \times \frac{N}{\prod_{i=0}^r q_i}~.
	 \label{eq:Pi_parametrized}
\end{equation}
which constitutes our main object of study. 
Given a population of $N$ individuals, for a given set of parameters $\kappa, \omega, \eta$ and $\beta$, 
our goal is to determine the optimal hierarchical structure, characterised by its
number of hierarchical levels $p^*$ and the associated group sizes $\{q_0, \ldots, q_{p^*}\}$,
which maximise the production \eqref{eq:Pi_parametrized}.

\subsection{Military hierarchies}

Instead of \eqref{eq:coef_definition}, it also instructive to study the special case $\mu_r=  \omega \delta_{r0}$ while keeping $\lambda_r = \rho^r$.
This represents the situation where only the ground level produces actual output,
whereas the higher order levels are only acting as coordination nodes, for instance to allocate resources, manage and control. 
We shall refer to this as the \textit{military hierarchy}, in reference to the fact that it is often the lowest military ranks 
(starting with ``privates'') who are exposed to active combats (new technology may be changing this),
and the higher levels mostly exert ``command and control''. 

This case with the approximation $q(q-1) \approx q^2$ allows for an analytical treatment given in the SI. 
As an illustration, with $N = 2^{12} = 4096, \beta=1.5, \rho = 0.5, \omega=6$,
the optimal production is $\Pi = 40'447$ with $p^*=4$ and the optimal structure is given by
$(q_0 \approx 9.1,q_1 \approx2.8,q_2 \approx2.8,q_3 \approx 6.5,q_4 \approx 8.9)$.

Varying $\omega$, we find a qualitative differences between the 
hierarchical structures for large versus low $\omega$'s, which illustrate the fight between having large groups
at the bottom to enhance productivity and the cost of coordination: for small $\omega$'s,  
the optimal structure consists in having maximally fragmented hierarchical structures with a maximum number of levels
and smallest group sizes (all $q_r$'s are equal to the minimum size $2$); for larger $\omega$'s,
larger subgroups are favoured, especially at the bottom and top levels, with relatively fewer levels
(see Fig. 1 in the SI)

\section{Optimal structures for geometrical hierarchies} 
\label{sec:special_case}

\subsection{Description of optimal structures}
\label{sec:opt_structure}

\begin{figure*}[!htb]
	\centering
	\includegraphics[width=\textwidth]{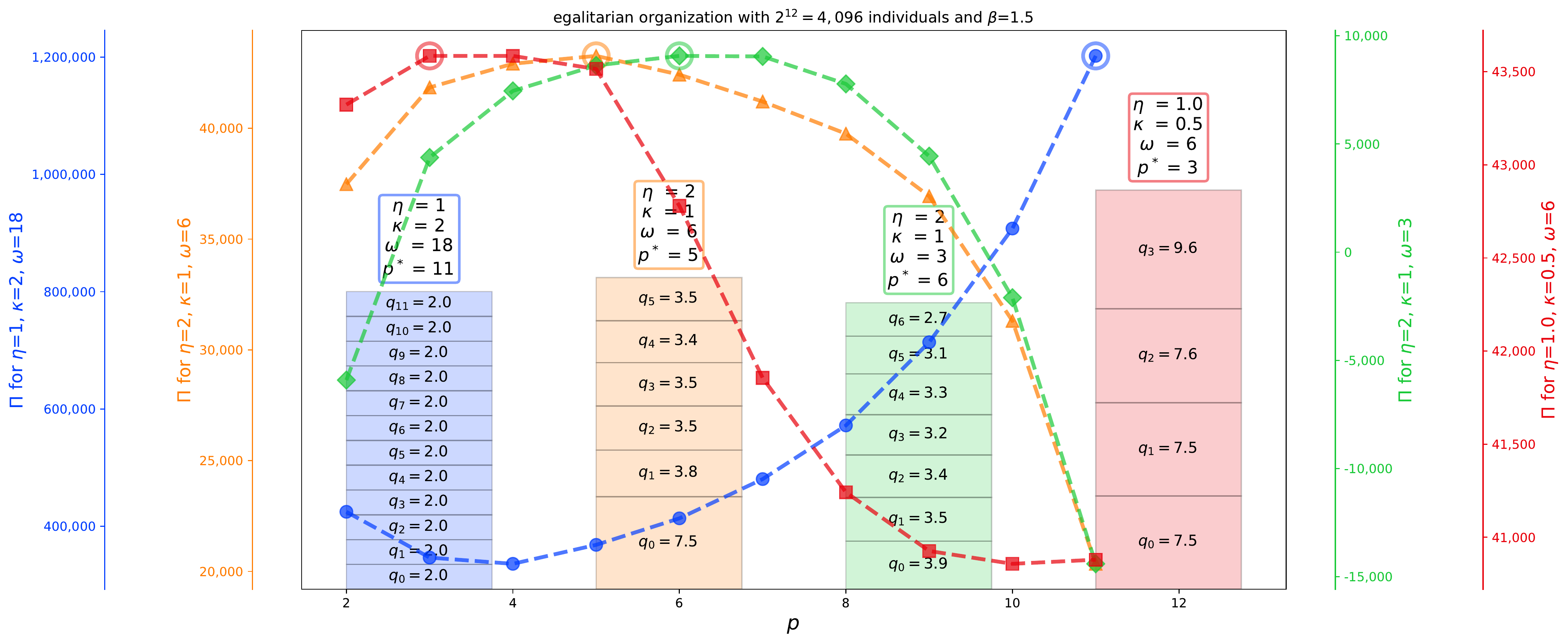} 
	\caption{
			Production $\Pi(p)$ as a function of $p$ (number of hierarchical levels minus one) for four different sets of parameters
			for a total population of $2^{12}=4096$ collaborators.
			For each of the four parameter set indicated in the four legends along
			the vertical axis, we obtain the optimal group sizes $\{q_0, \ldots, q_{p}\}$, and calculate the corresponding 
			total production $\Pi(p)$ from expression \eqref{eq:Pi_parametrized}. The function $\Pi(p)$ exhibits
			a maximum at some $p=p^*$ indicated with an open circle. 
			For each of the four optima, the corresponding optimal group sizes $\{q_0, \ldots, q_{p^*}\}$ are given in the 
			form of a stack of rectangles put on top of each other.
			The four different sets of parameters span different regimes and thus hierarchical designs.  Non-integer values of
			$q_r$'s should be interpreted as a combination of group of integer numbers of collaborators, with numbers within
			one unit from the quoted $q_r$ and such their average value is as close as possible to the $q_r$. For instance, 
			$q_r=3.7$ or $3.8$ should be interpreted as corresponding to three groups of $4$ and one group of $3$.
			See main text for a detailed description of the four different cases. 
			}
	\label{fig:p_dependence}
\end{figure*}

We now analyse the configurations of group sizes $(q_0, \ldots, q_p^*)$ that maximise \eqref{eq:Pi_parametrized}.
For a hierarchy with some fixed $p$, we determine the configuration $\{q_0, \ldots, q_p\}$ iteratively, 
by solving equation $\partial \Pi \left/ \partial N_r \right. =0$, 
and verifying that the solution indeed corresponds to a maximum.
Details are found in the SI. There, we also double check that the optimal solution is not obtained 
by splitting the $N$ individuals into isolated sub-structures.

The optimal group sizes are found to obey the recursive relation
\begin{equation}
	\mu_{r+1} \beta q_{r+1}^{\beta-1}  + 2 \lambda_{r+1} q_{r+1}  = \mu_r (\beta-1) q_r^\beta - \lambda_r q_r^2 - \lambda_{r+1}, 
	\label{eq:q_r__optimal}
\end{equation}
in the presence of the constraints $N = \prod_{r=0}^p q_r$ and $q_r \in (2, N / 2^p)$. 
If solutions of \eqref{eq:q_r__optimal} violates these constraints, one has to consider solutions on the boundaries. 
We thus apply a sequential numerical optimisation. 
First, for each $p \in \{0,1, \ldots, p_\text{max} \}$, we obtain numerically the
configuration $\{q_0, \ldots, q_p\}$ that maximises \eqref{eq:Pi_parametrized} (see SI for details),
thus getting the total production $\Pi(p)$ as a function of $p$.
Different examples of this $p$-dependence are depicted in Fig. \ref{fig:p_dependence}, 
Then, $p^*$ is determined as the value that maximises the total production $\Pi(p)$ given by \eqref{eq:Pi_parametrized}.
Fig. \ref{fig:p_dependence} presents the results of the search for the optimal hierarchical structures 
for four different set of parameters.

\noindent
{\it (a) Small group sizes and many hierarchical levels}: An end-member class of solutions consists in having the smallest groups as possible, 
structured over as many levels as possible. This corresponds to the hierarchical structure
at the boundary of the constraints $q_r \in (2, N / 2^p)$, namely $q_r=2$ at all levels $r$ and thus $N=2^{p+1}$.
This occurs approximately (but not necessarily precisely) when the output pre-factor $\omega$ is large compared to the
relative cost-productivity scaling factor $\eta$ and groups at higher levels are equally or more efficient 
than lower levels ($\kappa \geqslant 1$). This generalises the results found for the military hierarchies, whose structures
are more simply controlled by the production $\omega$ of the bottom level with an inverse dependence as
a function of $\omega$, illustrating that hierarchical structures result from subtle competition
between the different ingredients $\omega, \kappa, \eta$.

\noindent
{\it (b) Trade-off solutions with non-trivial group sizes at different levels}:
When the output factor $\omega \kappa^r$ and cost factor $\rho^r :=(\kappa \eta)^r$ are more balanced over multiple levels of the hierarchy, 
solutions expressing a trade-off between output and cost are characterised by
non-trivial optimal group sizes at different levels of the hierarchy. This is illustrated in Fig. \ref{fig:p_dependence}
by the orange line with filled triangles and the stack of orange rectangles giving $p^*=5$ and the corresponding optimal group sizes.


\noindent
{\it (c) Decreasing production with hierarchical level}:
For productions that decay with level order $(\kappa < 1)$, a small number of hierarchical levels is preferred ($p^*=3$), 
which can, for instance, be combined with group sizes
that are increasing at higher orders in the hierarchy (red line with filled squares in Fig. \ref{fig:p_dependence}). 
Small optimal values for $p^*$ are also found analytically for the ``military hierarchy'', which is an extreme case
of decrease of production with level order (see SI).


\subsection{Dependence of hierarchical structure properties as a function of population size}

\begin{figure}[!htb] 
	\centering
	\includegraphics[width=0.5\textwidth]{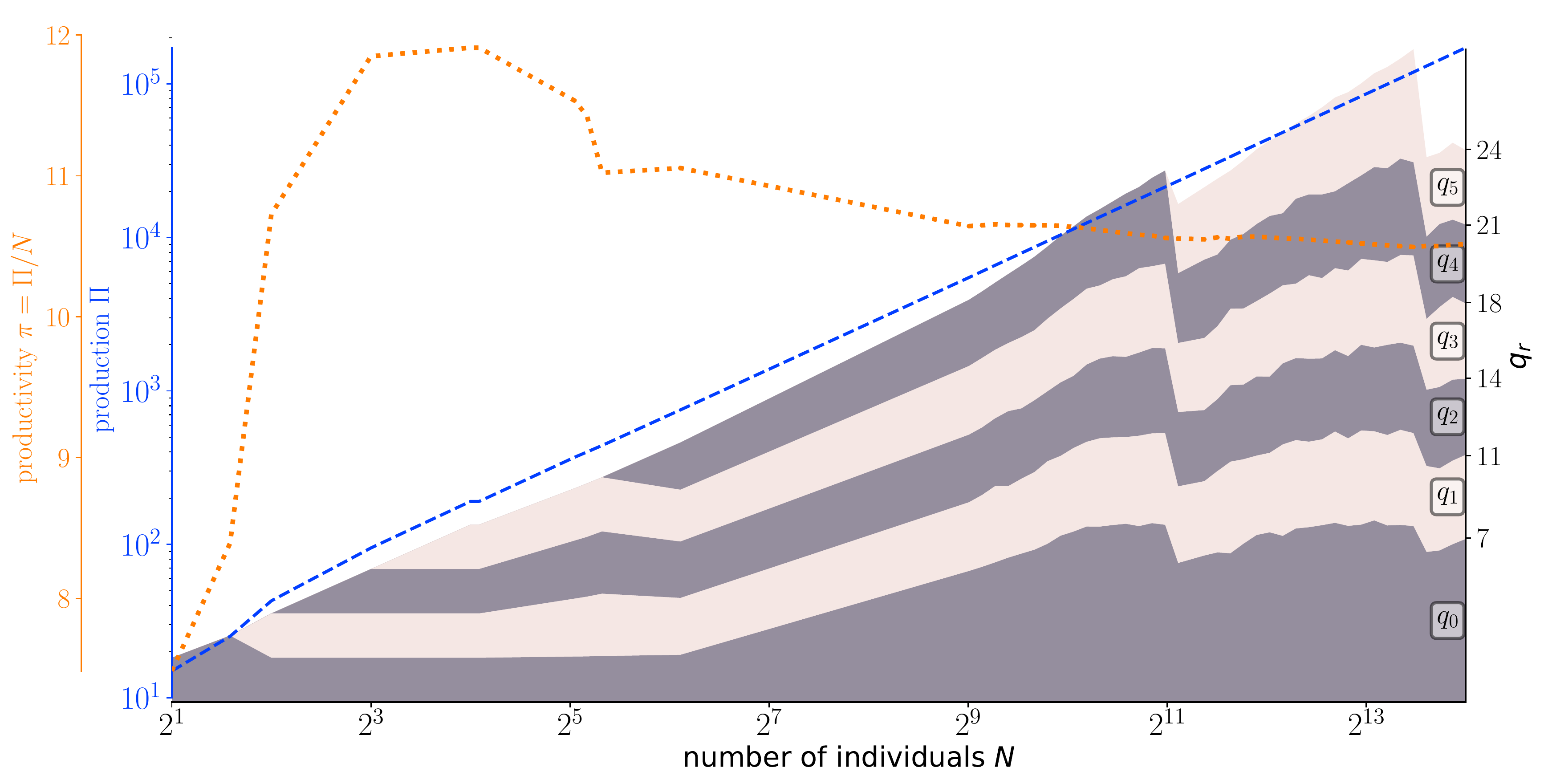} 
	\caption{For  $\beta = 1.5, \kappa = 1, \omega = 6$ and $\eta = 2$, we show the dependence as a function of
	the total population number $N$ (in logarithmic scale expressed in powers of $2$) 
	of three variables characterising the optimal hierarchical organisation
	determined and described in subsection \ref{sec:opt_structure}: (i) total production $\Pi$ (blue dashed line);
	(ii) productivity per individual, $\pi \equiv \Pi / N$ (dotted orange line); (iii) optimal group ratios $q_r$ 
	shown in stacked bands of alternating colours of dark grey and pink. 
	The tick marks on the right $y$-axis show the sizes $q_r$ of the groups for $N = 2^{14}$, for which the optimal structure is given by 
	$p^* = 4$ with $q_0 \approx 7, q_1 \approx 4, q_2 \approx 3, q_3 \approx 4$, $q_4 \approx 3$ and $q_5 \approx 3$.} 
	\label{fig:N_dependence}
\end{figure}

For given productivity characteristics $\{\beta, \omega, \kappa\}$ and coordination cost properties $\{\eta\}$ (or $\rho$)
corresponding to case (b) above with $p^*=5$ in Fig. \ref{fig:p_dependence}, 
Fig. \ref{fig:N_dependence} shows the dependence of three main features of the optimal hierarchical structure as a function of the population size $N$. 
The optimal group ratios $q_r$ shown in stacked bands of alternating colours of dark grey and pink
exhibit several interesting features. First, as $N$ increases, the optimal number $p^*$ of levels exhibit a series of 
transitions, from 
$p^*=0$ to $p^*=1$ at $N=3$, 
from $p^*=1$ to $p^*=2$ at $N=4$, 
from $p^*=4$ to $p^*=5$ at around $N=2020$, 
and so forth.
Note that the range of population sizes for a given $p^*$
does not follow a simple geometrical series that would be revealed by an approximately equi-spaced spacing
in the logarithmic representation of the x-axis in Fig. \ref{fig:N_dependence}. 
In particular, the optimal value $p^*=4$ is found over a very large
interval $2^5-1 < N \leq 2^{11}$. Nonetheless, $p^*$ can be shown to grow asymptotically
on average proportionally to $\ln N$ (see SI).

Each of the transitions in $p^*$ is mirrored by a break or spike in the 
dependence of the productivity per individual, $\pi \equiv \Pi / N$, as a function of $N$.
The first regime with $p^*=0$ corresponds to a super-linear growth of production, until it saturates
with the emergence of the second hierarchical level, which is needed to tame the growing cost of communications.
In particular, $\pi$ has its absolute maximum at $N \approx 15$, suggesting an optimal size of $15$ for an independent organisation,
which should be organised into $q_1=2$ teams of $q_0=7$ members. We stress that these numbers are 
the optimal ones for the specific parameters $\beta = 1.5, \kappa = 1, \omega = 6$ and $\eta = 2$.
Other parameters would lead to different optimal hierarchies. 
Last, the productivity $\pi$ can be seen to converge to $\pi_\infty \approx 10$ for large $N$,
corresponding to an asymptotically linear increase of the total production $\Pi$ as a function of organisation size $N$.
As $\pi_\infty \approx 10 > \omega=6$, the production per capita in the optimal hierarchical organisation is approximately 
$67\%$ larger than that of isolated individuals, giving a significant gain. This asymptotic productivity
per individual is however about $33\%$ smaller than that of the optimal population size $N \approx 15$,
exemplifying the relative disadvantage of growing organisations even with its optimal hierarchical structure.
In a flat organisation,  the quadratic cost would always end up dominating the total production and lead to a collapse of the organisation.
Only a hierarchical structure can relieve from the excruciating cost of coordination and harvest the superlinear 
productivity ($\beta >1$).	The overall lesson is that knowledge
of production and cost properties should provide guidance to shape the organisation structure for better productivity
and performance. This has implications, not only for growing organisations that should develop additional levels
of hierarchy in stage, as illustrated in Fig. \ref{fig:N_dependence} but also, for mergers and acquisitions. 

\subsection{When is the whole more than the sum of the parts?}
\label{sec:phases}

As mentioned above and shown more systematically in the SI, the total production $\Pi^*(N) = \pi(N) N$
of the optimal hierarchical organisation, for any fixed set of productivity and cost parameters
such that $\pi(N)>0$, scales asymptotically linearly with the number of individuals $N$. In other words, the productivity
or production per capita $\pi(N)$ converges to a constant $\pi_\infty$ for large $N$, which is a function 
of $\beta, \omega, \kappa$ and $\eta$.
On the other hand, for $N$ non-interacting individuals (i.e. for $N$ ``structures'' of $q_0 = 1$
individual each), equation \eqref{eq:Pi_parametrized} reduces to a total production $\Pi^I(N) = \omega N$,
i.e., $N$ times the production $\omega$ of a typical individual. 
In competitive, free markets, it will be rational for people to come together and cooperate only if their per 
capita production turns out to be larger than their individual ones.

\begin{figure}[!htb]
	\centering
	\includegraphics[width=0.5\textwidth]{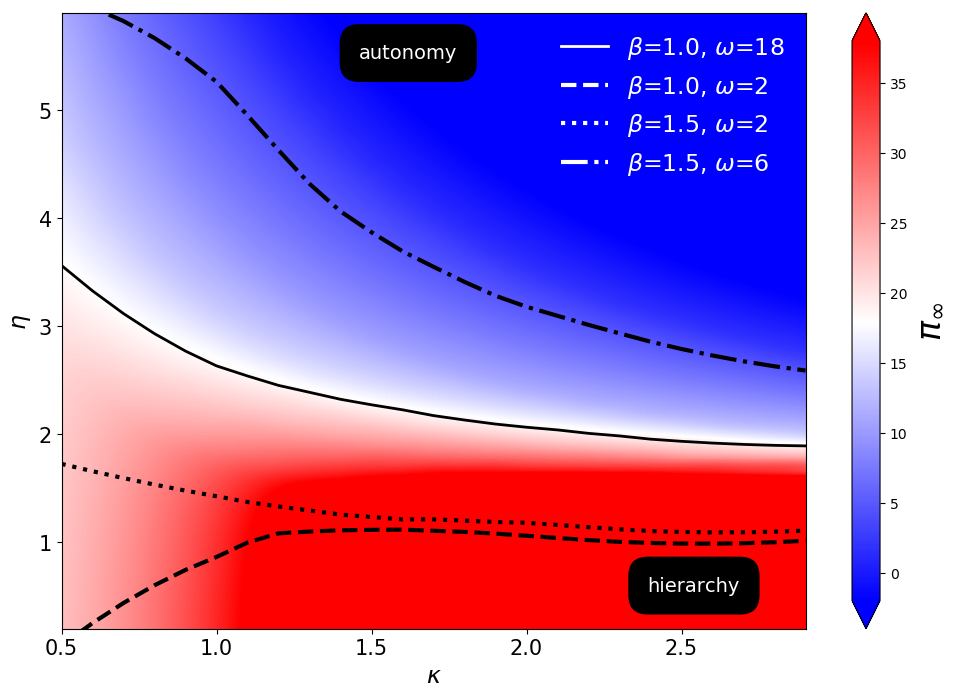}
	\caption{	For fixed values of $(\beta, \omega)$ given in the inset, the domain below each curve is such that the production 
			per capita, $\pi_\infty$, in the optimal hierarchical structure is larger than the production $\omega$ of an isolated individual. 
			The computation of the production per capita has been performed numerically for $N = 2^{14}$, which is large enough that the 	productivity per individual has approximately converged to its asympotic value $\pi_\infty$.
			The background colouring shows $\pi_\infty$ for $\omega=18, ~\beta=1$. 
			The change of regime (white background) is where $\pi_\infty = \omega$, i.e. exactly where the hierarchical output $\Pi$ is equal to the 
			input of $N$ individuals, i.e. where $\Pi(N) =  \omega N$. 
			For parameters ($\kappa,\eta$) above this line, it is more productive if the $N$ individuals work on their own. 
			Below this line, hierarchical organisation is preferred. }
	\label{fig:regimes}
\end{figure}

Fig. \ref{fig:regimes} delineates the domain in the $(\kappa, \eta)$-plane, for different sets of fixed values of $(\beta, \omega)$. 
The domain can be split into two regimes, one for which $\pi_\infty \leq \omega$ (regime where individuals are better off producing on their own, called ``autonomy''), 
and the complementary domain for which $\pi_\infty > \omega$ (regime where individuals are better off forming a group,  called ``hierarchy''). 
The curve separating the two domains is an increasing function of $\eta$ as a function of $\omega$. 
Intuitively, the larger the individual productivity $\omega$, the larger can be the relative cost-productivity scaling factor $\eta$ while still ensuring that a hierarchical society emerges.

The regime $\pi_\infty < \omega$ represents organisations whose goals are not necessarily
to improve productivity but to be stronger than other polities as a whole. Indeed, 
for many societies engaged in military competition for instance,  
what matters is the total military power relative to its rivals, not per soldier efficiency. 
Our theory on optimal hierarchical organisations applies there as well, as we obtain 
non-trivial hierarchical organisations even for cases where $\beta \leqslant 1$. 
These solutions are mostly dominated by a minimisation of the communication overhead.
Elaborations of this regime will be reported elsewhere.

\section{Triadic hierarchies \& optimal span of control} 
\label{sec:triadic}

\begin{figure*}[!htb]
	\centering
	\includegraphics[width=\textwidth]{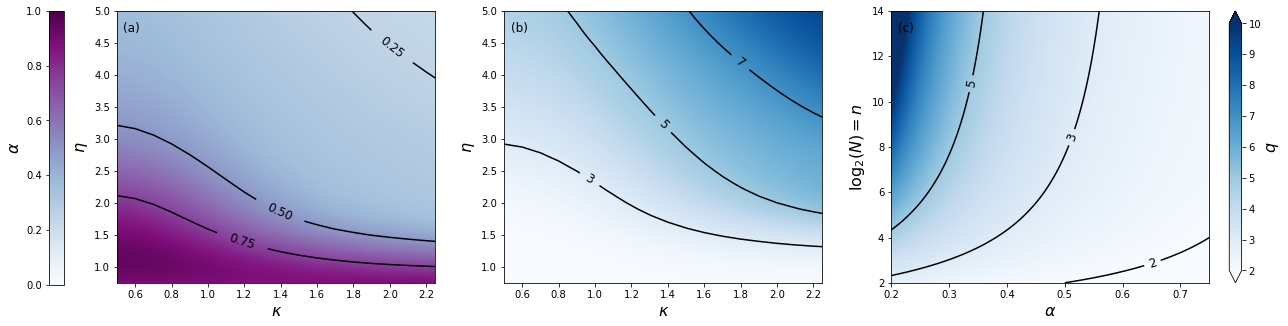}
	\caption{	(a) Estimated asymptotic coefficient $\alpha$ of the linear relation $p^* = \alpha \cdot (n-1)$, where $n = \log_2 N$, for different parameters $\kappa, \eta$ and fixed $\beta=3/2, \omega=2$. 
				We observe that $\alpha$ varies from its minimum value close to zero all the way up to its maximum value of one.
			(b) Same as figure (a), but instead of $\alpha$, we map the optimal scaling ratio $q$ via \eqref{eq:q_triadic}, with $n = 12$, i.e. for $N = 2^{12} = 4096$ individuals.
			(c) Optimal scaling ratio \eqref{eq:q_triadic} as a function of $\alpha$ and (logarithm of) number $n$ of individuals. 
				For a wide set of parameters, the scaling ratio is found within the empirically observed range from 2-4.
			 } 
	\label{fig:alpha}
\end{figure*}

In section I-\ref{sec:Dunbar_number}, we suggested a derivation of Dunbar's number $\sim 100-250$, describing the maximum number of 
people with whom one can develop stable social relationships. But Dunbar's number is actually just a part of the full
story. In 2005, Zhou et al.\cite{Zhou2005} discovered the general existence of a 
discrete hierarchy of group sizes with a preferred scaling ratio close to three:
humans spontaneously form groups of preferred sizes organized in a geometrical
series approximating $3-5, 9-15, 30-45, 90-140, 250-400$ and so on. This finding has been 
corroborated in many different contexts 
\cite{Fuchs2014,Dunbar2015,Dunbar2018,CarronKaskiDunbar2016} as well as for various groups of animals \cite{RussellDunbar2008}.
These works quantify the qualitative 
anthropological studies showing that societies, from primates \cite{Dunbar2018} to humans \cite{Kottak2011},
tend to arrange into discrete hierarchical structures, 
with group sizes ratios between hierarchical levels that typically range from $2$ to $4$  \cite{Zhou2005}.  
Within our framework embodied in equation \eqref{eq:Pi_general_explicit}, this observation finds a natural explanation, as we will now show. 

As long as the coefficients in the sets $\{\mu_r\}$ and $\{\lambda_r\}$ ensure that a group (which has to be hierarchical) is more optimal than $N$ isolated individuals,
the optimal number of hierarchical levels $p^*$ scales  as $p^* \sim \log_2 N  -1 = n-1$, where we define $N = 2^n$ for convenience (cf. SI). 
Since the maximum number of hierarchical levels is given by $p_\text{max} = n-1$, 
which occurs when all scaling ratios are equal to the minimum $q_r= 2$, 
one can deduce that $p^* = \alpha \cdot (n-1)$ for some $\alpha \in (0,1)$. 
So for example, in Fig. \ref{fig:N_dependence}, we see that the asymptotic regime ($\pi$ is constant) starts roughly around $N=2^{5}$, at the beginning of the $p^*=4$ layer. 
The $p^*=5$ layer then only occurs at $N=2^{11}$, such that we estimate $\alpha \approx (5-4) / ( (11-1) - (5-1) ) \approx 0.17$. 
A more robust way to estimate $\alpha$ is outlined in the SI, 
and a systematic classification of $\alpha$ as a function of different parameter configurations, 
$\alpha = \alpha(\beta,\eta,\omega,\kappa)$, is depicted in Fig. \ref{fig:alpha} (a), 
showing that $\alpha$ ranges from its the minimal value close to zero all the way up to its maximum at one. 

Assuming for simplicity that $q_r \approx q_s ~  \forall r, s$, it follows that
$2^n = N = \prod_{r=0}^{p^*} q_r \approx q^{p^* + 1} = q^{\alpha n + 1}$ 
and hence 
\begin{equation}
	q \approx 2^{\frac{1}{\alpha + 1/n}}. 
	\label{eq:q_triadic}
\end{equation}
We can thus map the coefficient $\alpha$ to an optimal scaling ratio $q$ through \eqref{eq:q_triadic}. 
The optimal scaling ratio $q$ is depicted in Fig. \ref{fig:alpha} (b) for different sets of parameters, showing that $q \sim 2-4$ holds over a wide range of parameters.  This is further exemplified in \ref{fig:alpha} (c), which plots the functional relation \eqref{eq:q_triadic} for a large range of values of both $\alpha$ and $n$. 

However, there are other regimes where $q$ deviates significantly from the range $2-4$, and depends on the level
within the hierarchy. We propose that this range of parameters and corresponding regimes
explain the findings in Business Management on the span-of-control \cite{Graicunas_Span33,Entwisle_Span61,Tarng_Span88}, 
which is concerned with
 the number of subordinates a supervisor can or should manage. In many Fortune 500 organisation, the so-called ``hourglass'' organisation
 is observed, characterised by the vice-presidents at the top presiding over 8 to 9 senior directors, each of the senior 
 director controlling 6 to 8 directors, each director supervising 3 to 6 lead managers, each lead manager directing
 4 to 6 managers, each manager overseeing 5 to 7 supervisors, each supervisor leading 8 to 14 employees \cite{Buchanan_Span03}.
 Such a structure is strongly reminiscent of the optimal hierarchy shown in Fig. 1 of the SI for 
the ``military'' organisation with
$\beta=1.5, \rho=0.5$ for large production per individual ($\omega=6$ or $10$). 
We thus find that organisations, where the production is essentially done at the bottom level 
and for which the other higher levels are only present to optimise coordination and control, are characterised
by strong non-universal scaling ratios that are level-dependent, with span-of-control ranging from $3$ to $20$.
In contrast, as shown above, when the production is more evenly contributed by all levels, 
a quasi-universal scaling ratio in the range $2-4$ ensures the optimal functioning of the society.

In conclusion, we have shown that, with very few ingredients captured in equation \eqref{eq:Pi_parametrized}, 
a wide variety of hierarchically ordered complex organisational structures can be derived. 
Future works will include pure integer optimisation, 
whereby we keep the group sizes equal to integer values, while simultaneously allowing for different group sizes on the same hierarchical level.
Our fractional group ratios $q_r$ can then be seen as averages over these group sizes. 
Such an integer optimisation allows for direct comparison with actual organizational structures. 
Other extensions include allowing for heterogeneity among individuals in the productive ability, 
complementarity and different communication and coordination cost functions.

\section*{Acknowledgements} 

We are thankful to R. Dunbar and an unknown referee for stimulating inputs. 
SC Lera acknowledges support from the SNF grant Early.PostDoc.Mobility.

\section*{References}
\balance
\bibliographystyle{unsrt}
\bibliography{bibliography}

\end{document}